\documentstyle[amssymb,prb,aps,epsf,floats]{revtex}

\newcommand{\ie}{\mbox{i.\ e.\ }}
\newcommand{\eg}{\mbox{e.\ g.\ }}

\newcommand{\etal}{\mbox{\it et.\ al.\ }}
\newcommand{\be}{\begin{eqnarray}}
\newcommand{\en}{\end{eqnarray}}
\newcommand{\no}{\nonumber}

\newcommand{\hc}{\mbox{\it h.\ c.\ }}
\newcommand{\mc}{\mathcal}
\newcommand{\up}{u}
\newcommand{\dn}{d}
\newcommand{\upd}{u^{\dagger}}
\newcommand{\dnd}{d^{\dagger}}
\newcommand{\f}{f}
\newcommand{\p}{p}
\newcommand{\fd}{f^{\dagger}}
\newcommand{\pd}{p^{\dagger}}
\newcommand{\at}{\tilde{a}}
\newcommand{\atd}{\tilde{a}^{\dagger}}

\begin{document}


\title{
A novel spin wave expansion, finite temperature 
corrections and order from disorder effects in 
the double exchange model.
}

\author{Nic Shannon}
\address{
Max--Planck--Institut f{\"u}r Physik komplexer Systeme,
N{\"o}thnitzer Str. 38, 01187 Dresden, Germany.
}
\author{Andrey V. Chubukov}
\address{
Department of Physics, University of Wisconsin--Madison,
1150 Univ. Av., Madison WI 53706--1390, USA.
}
\date{\today}
\maketitle

\begin{abstract}
The magnetic excitations of the double exchange (DE) model are 
usually discussed in terms of an equivalent ferromagnetic Heisenberg 
model.
We argue that this equivalence is valid {\it only} at a 
quasi--classical
level --- both quantum and thermal corrections to the magnetic 
properties of DE model differ from {\it any} effective Heisenberg 
model because its spin excitations interact only indirectly, through 
the exchange of charge fluctuations. 
To demonstrate this, we perform a novel large $S$ expansion for the
coupled spin and charge degrees of freedom of the DE model, aimed at 
projecting out all electrons not locally aligned with core spins. 
We generalized the Holstein--Primakoff transformation to the case 
when the length of the spin is by itself an operator, and explicitly 
constructed new  fermionic and bosonic operators to fourth order 
in $1/\sqrt{S}$. 
This procedure removes all spin variables from the Hund coupling
term, and yields an effective Hamiltonian with an overall scale of
electron hopping, for which we evaluate corrections to the
magnetic and electronic properties in $1/S$ expansion to order 
${\mc O}(1/S^2)$.  We also consider the effect of a direct 
superexchange antiferromagnetic interaction between core spins. 
We find that the competition between ferromagnetic double 
exchange and an antiferromagnetic superexchange  provides a new 
example of an "order from disorder" phenomenon ---  
when the two interactions are of comparable strength, 
an intermediate spin configuration (either a canted or a spiral state) 
is selected by  quantum and/or thermal fluctuations.  
\end{abstract}



\section{Introduction}

Many magnetic systems of great experimental interest, for example 
the colossal magnetoresistance (CMR) Manganites \cite{xener,anderson} 
and Pyroclhores \cite{pyrochlores}, comprise a band of itinerant 
electrons interacting with an ordered array of localized magnetic 
moments with spin $S$.
In many cases these systems can be modeled as a single tight band of 
electrons interacting with localized core spins by a Hund 
rule exchange interaction, and can be described by a Hamiltonian 
of the form :
\be
{\mc H} &=& {\mc H}_0 + {\mc H}_1 + {\mc H}_2
\label{eqn:KondoH}
\en
where ${\mc H}_0$ is the bare Hamiltonian for itinerant electrons
\be
\label{eqn:H0}
{\mc H}_0 &=& - t \sum_{\langle ij \rangle\alpha} 
   c^{\dagger}_{i\alpha} c_{j\alpha},
\en
${\mc H}_1$ describes the first Hund rule (Kondo type) coupling 
between localized spins and itinerant electrons
\be
{\mc H}_1 = -J_H \sum_{i} \vec{S}_i.\vec{s}_i  
   &\qquad&
\vec{s}_i = \sum_{\alpha\beta}       
   c^{\dagger}_{i\alpha}
   \vec{\sigma_{\alpha\beta}}c_{i\beta}
\en
and ${\mc H}_2$ describes a superexchange between 
localized spins 
\be
{\mc H}_2 &=& - J_2 \sum_{\langle ij \rangle} \vec{S}_i \vec{S}_j
\en
Here the indices $i$ and $j$ run over lattice sites, 
$\alpha$ and $\beta$ over electron spin states,
and components of $\vec{\sigma}_{\alpha\beta}$ are Pauli matrices.

Our primary interest will be to study the limit in which the Hund's 
rule coupling $J_H$ is positive and is much larger than both $J_2$ 
and $t$. 
In this case itinerant conduction electrons must be locally aligned 
with the core spin on any site. As the kinetic energy of electrons 
is minimal when all electrons are parallel, the core spins are also 
all parallel, \ie the ground state is a ferromagnet.
The ferromagnetic interaction between core spins mediated by 
conduction electrons is often referred  to as ``double exchange'', 
and the model described by ${\mc H}_0 +{\mc H}_1$, in the limit 
$t/J_H \to 0$, is often referred to as the double exchange 
ferromagnet (DEFM). 

Many authors have demonstrated that for classical spins,
the fermion-mediated ferromagnetism is described by the effective
nearest-neighbor Heisenberg model
\be
\label{eq:HFM}
-J_1 \sum_{\langle ij \rangle} \vec{T}_i. \vec{T}_j
\label{exchange}
\en
with $T = S+x/2$, where $x$ is the electron density, and 
ferromagnetic exchange integral
\be
J_1 = \frac{\overline{t}}{2S^2}
\en
where $\overline{t}$ is the kinetic energy per bond in 
the lattice, which is a function of the electron doping $x$.   
In the limit of low electron density $x \to 0$, 
$\overline{t} \sim tx$.
 
The aims of this paper are two--fold.  
Firstly, we discuss the extent to which the DEFM is equivalent 
to the nearest--neighbor Heisenberg ferromagnet. 
We argue that this equivalence holds only at infinite $S$, when spin 
waves are non--interacting quasiparticles.  We demonstrate that the 
interaction between spin waves in the DEFM is qualitatively different 
from that in the Heisenberg ferromagnet, for which spin waves are exact
eigenstates of the Hamiltonian.
This difference gives rise to different forms of quantum and thermal 
corrections to the spin wave dispersion.  
In particular, we show that spin excitations in the DEFM have a 
finite lifetime even at $T=0$, \ie that they cannot be true 
eigenstates of the Hamiltonian.
Physically, the difference between the two models stems from the 
fact that in a DEFM, the dynamics of the bosonic spin wave modes
are completely determined by those of the itinerant electrons.
Just as the existence of a finite density of charge carriers 
generates a dispersion for spin waves about the groundstate of 
the DEFM, so fluctuations of charge density generate a retarded 
interaction between these spin waves.
Since this interaction is proportional to the charge susceptibility 
of the itinerant electrons, both its value and 
the dependence on momentum and frequency are very different from that 
in the Heisenberg model.

\begin{figure}[tb]
\begin{center}
\leavevmode
\epsfysize 6cm
\epsffile{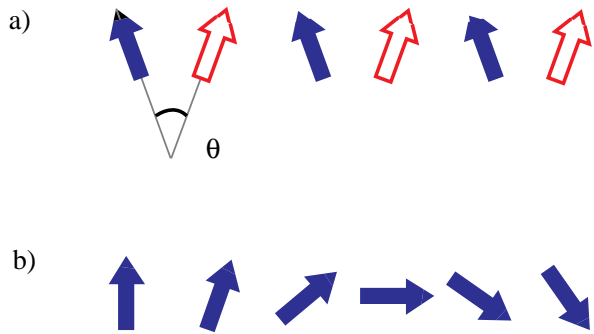}
\caption{
Two sublattice canted (a) and spiral (b) spin configurations.
}
\label{canting}
\end{center}
\end{figure}

Our second goal is to analyze what happens when the ferromagnetic DE 
interaction competes with the antiferromagnetic superexchange. 
The trade--off between fermion-mediated and direct exchanges
can, in principle,  lead to many different ground states,
and indeed numerical studies suggest a very rich phase diagram
\cite{numerical}.
Here we address the issue how the system evolves from a DEFM to and 
AFM with increasing $J_2$, provided that there is no phase 
separation. 
As observed by de Gennes \cite{degennes}, for classical spins, the
configurations which interpolate between FM and AF order are the ones 
in which the neighbouring lattice spins are misaligned by an angle 
$\theta$ such that $\cos \theta/2 = J_1/J_2$.  This criterium can be 
satisfied by canting the spins into a two-sublattice  state shown in
Fig.~\ref{canting}a or into a spiral state shown in Fig.~\ref{canting}b.
However, these two configuration are not the only possible as at the
classical level we may take any spin of, say, $B$ sublattice of the 
canted phase and rotate it about the direction of magnetization of 
the $A$ sublattice without altering the angle between it and the 
neighbouring spins. This rotation introduces an infinite
set of classically degenerate intermediate configurations which 
interpolate between canted and spiral states. 
We performed a spin wave analysis of the canted phase using the 
transformation to bosons described below, and indeed found that 
a local degeneracy yields a branch of spin wave excitations 
with  $\omega_{sf} (q) =0$ for all $q$.
Since it costs no energy to make an excitation, the system cannot 
distinguish between different states, and is  magnetically
disordered even at $T=0$\cite{degennes,golosov1}.

This argument, however, does not hold for quantum spins, and we 
expect that quantum fluctuations enable the system to choose its
true groundstate. 
A widely studied example of this ``order from disorder'' behavior
is provided by the highly frustrated 2D antiferromagnet on a 
Kagom\'e lattice \cite{andrey}.  The question remains as to what 
kind of order is preferred in our case.
  
To address this issue we analyzed what is the momentum of the 
instability of a FM configuration. 
For classical spins, as discussed above, DE ferromagnetism 
can be described by the effective exchange  model Eq.~(\ref{exchange}). 
In this situation, the excitation spectrum in a FM phase is simply
\be
\label{eqn:dd}
\omega_{sw}(q) &=& zS  (J_1 - J_2) (1 - \gamma_q) 
\en
where $z$ is the lattice coordination number, and
\be 
\gamma_q &=& \frac{1}{z} \sum_\delta e^{i \vec{q}.\vec{\delta}}
\en
where the sum on $\{\vec{\delta}\}$ runs over nearest neighbors
vectors.
This excitation spectrum vanishes identically for $J_1 = J_2$, in 
agreement with the infinite degeneracy at $J_1<J_2$.
Suppose now that this degeneracy is lifted by quantum fluctuations,
and the spin wave spectrum first becomes unstable at some momentum 
$Q$.
One can easily demonstrate (see below) that if $Q=0$, the resulting 
state is a spiral, if $Q=(\pi,\pi,\pi)$, the instability leads to 
a canted state, and if $Q$ is in between the two limits, the system 
chooses some intermediate spin configuration.
 
We show that for both small and large electronic densities
($x \ll 1$ or $1-x \ll 1$), the DEFM becomes unstable 
(with increasing $J_2$) against the two--sublattice canted structure. 
At intermediate densities ($0.05 < x < 0.68$ in 3D), the first 
instability is against a spiral spin configuration.  
Similar results hold for the $2D$ case.
We also show that thermal corrections to $\omega_{sw}(q)$ 
are very different from those in the Heisenberg ferromagnet, and in 
3D can be even of a different sign at the lowest densities of 
carriers. 
For realistic densities, we found that the functional form of the 
thermal correction is approximately the same as in the Heisenberg 
model, but the overall amplitude is substantially increased.

To proceed with the spin wave calculations, we need a 
``bosonization'' scheme which treats both core spins and itinerant 
electrons on equal footing.  Our approach is to transcribe the model 
Hamiltonian of Eq.~(\ref{eqn:KondoH}) in terms of the 
``natural'' collective coordinates of the Kondo coupling term under 
the assumption that the size of the core  spin $S \gg 1$.   
This procedure has the advantage of making 
a clean distinction between different types of excitations (spin and 
charge modes) and of showing clearly the separation of different 
physical energy scales in the atomic limit.  
The transformation and its derivation are presented in the next Sec.
\ref{transformation}.
In Sec. \ref{DEFM}, we apply this transformation to the DE model and 
discuss the form of the thermal and quantum corrections to the 
spin wave spectrum.  
In this section, we also present the results for 
the damping rate of spin excitations.  In Sec. \ref{orderfrom}, we 
discuss how quantum and/or thermal fluctuations select 
an intermediate configuration which interpolates between 
ferromagnetic state at $J_1 >J_2$ and antiferromagnetic state at 
vanishing $J_1/J_2$. We also discuss here a peculiar re-entrant 
transition between spiral and canted states, which is due to a 
competition between the selection by thermal and quantum 
fluctuations. 
Finally, in Sec. \ref{conclusions} we present our conclusions.

\section{Derivation of transformation}
\label{transformation}

Since we wish to work in the limit where
the Hund's rule coupling $J_H \gg t$ is the largest energy scale 
in the problem, it makes sense to treat the onsite Kondo--coupling
exactly, and to introduce hopping between sites a perturbation.
In the limit where $J_H \to \infty$, we can go one step further
and project out all
electrons not locally aligned with the core spins.
We should also define spin wave excitations such that they are
the true Goldstone modes of the order parameter, \ie transverse
fluctuations of the composite spin
\be
\vec{T}_i = {\vec S}_i
+ \frac{1}{2}\sum_{\alpha\beta} c^{\dagger}_{i\alpha}
{\vec \sigma}_{\alpha \beta} c_{i\beta}
\en
We can accomplish these goals by  
first introducing new Fermi operators $\fd_i$ and $\pd_i$ which 
create local
states with high ($T=S+1/2$) and low ($T=S-1/2$) 
total spin $T$, and then generalizing the Holstein--Primakoff
transformation to the case where the length of the spin $T$ 
is itself an operator, introducing a corresponding 
bosonic operator $\at_i$.  
Here we outline this procedure, which was introduced 
in \cite{us}.  A comparison with an exact solution
on two sites are given in \cite{nsps}.

Let us start by considering the ``Kondo Atom'' of a single localized 
spin and itinerant (spin degenerate) electron orbital 
\be 
\label{eqn:atom}
{\mc H}_1 = -J \vec{S}~\vec{s} 
\en  
We can diagonalize this Hamiltonian in by introducing the 
total spin operator 
\be 
\vec{T} &=& \vec{S} + \vec{s}
\en
such that  
\be 
{\mc H}_1 
   &=& -\frac{J}{2}
      \left[ \vec{T}^2 - \vec{S}^2 - \vec{s}^2 \right]\no\\
   &=& -\frac{J}{2}
      \left[ T(T+1) - S(S+1) - s(s+1) \right]
\en
Here the quantum number for the ``itinerant'' electron $\vec{s}$ 
can take on values $s=\{0,1/2\}$, and the total spin quantum number 
$T = \{S+1/2, S, S-1/2\}$, so we find three corresponding energy 
eigenvalues 
\be
\label{eqn:eigen}
\lambda &=& 
   \left\{ -\frac{JS}{2}, 0, 
   \frac{JS}{2}\left(1 + \frac{1}{S} \right)\right\}
\en
where the state with total spin $S$ and eigenvalue zero is doubly 
degenerate and corresponds to zero or double occupancy of the 
itinerant electron state.   

We can now define a new quantization axis for a composite spin $T$ 
and introduce the corresponding ``up'' ($\f$) and ``down'' ($\p$) 
fermionic states.  The magnitude of $T$ is then
\be
\label{eqn:newS}
T = S + \frac{\fd\f - \pd\p}{2}
\en
This procedure corresponds to making an $SU(2)$
rotation of the electron spin coordinates such that the new 
quantization axis follows the direction of the instantaneous 
local magnetization.
Accordingly, one approach to deriving an effective action for the 
DE model is to work in the $SU(2)$ rotated local frame and introduce 
a gauge field to describe the rotation of coordinates
for an electron hopping along any given bond of the lattice
(see \eg \cite{millis}).

For our purposes, it is however more advantageous to use a 
somewhat different procedure. 
First we note that to leading order in $1/S$, the states with 
$T = S \pm 1/2$ correspond to aligning or antialigning the
spin of a single itinerant electron with the localized spin on
that site.  At this level, we can replace the Kondo coupling 
with a local magnetic field 
and reduce ${\mc H}_1$ to  ${\mc H}_1 = \tilde{{\mc H}}_1 +const$
where  
\be
\label{eqn:zeem}
\tilde{{\mc H}}_1  
   &\approx& -\frac{JS}{2}\left[ \fd\f 
       - \pd\p \right]
\en
Second, we observe that we can 
obtain the correct eigenvalues of the full quantum mechanical 
problem by keeping (\ref{eqn:newS}) and modifying the 
simple ``Zeeman--splitting'' form of Eq.~(\ref{eqn:zeem}) to
\be
\label{eqn:tildeH1}
\tilde{{\mc H}}_1  
   &=& -\frac{JS}{2}\left[ \fd\f 
       - \pd\p\left(1 +\frac{1}{S}\right) 
       + \frac{\fd\f \pd\p}{S} \right]
\en
One can easily make sure that Eq.~(\ref{eqn:tildeH1}) yields 
correct eigenstates (\ref{eqn:eigen}) for any $S$.
The new local interaction 
\be
-\frac{J}{2}\fd\f \pd\p
\en
comes from the fact that the eigenvalues of 
${\mc H}_1$ for $T = S \pm 1/2$ are not symmetric 
about zero.   
Physically, it implies that if the system has 
both $\f$ and $\p$ electrons it can save energy for positive 
$J$ by putting them on the same site to male a total spin of 
length $T = S$.  

We see that if $J$ is the largest parameter in the problem
and $S$ is large (as we assume), the fermionic states created by 
the $p-$operator are separated in energy by $J(S+1/2)$
from fermionic states created by $f-$operator,and can be safely 
eliminated from the low--energy sector.  This is the obvious
advantage of introducing $f-$ and $p-$ operators. 
The problem which we now have to solve is how to relate these two 
operators, which describe fermionic states of a composite spin 
$T$ to the ``original'' up/down fermionic operators 
$c_{\uparrow} = \up$, $c_{\downarrow} = \dn$ introduced
with respect to a quantization axis of the core spin $S$.    
We solve this last problem order by order in $1/S$ by introducing 
two sets of Holstein--Primakoff bosons --- one $a$ for a core spin 
$\vec{S}$ and another $\at$ for the composite spin $\vec{T}$. 
Introducing the quantization axis $z$ for the core spin and
the corresponding fermionic states
$c_{\uparrow} = \up$, $c_{\downarrow} = \dn$, we have 
\be
S^z &=& S - a^{\dagger}a\\
S^+ &=& \sqrt{2S} \sqrt{1 - \frac{a^{\dagger}a}{2S}} a\\
S^- &=& 
   \sqrt{2S} a^{\dagger} \sqrt{1 - \frac{a^{\dagger}a}{2S}}
\en
and 
\be
T^z = &S^z + \frac{\upd \up - \dnd \dn}{2}&
  = T - \atd\at\\
T^+ = &S^+ + \upd\dn&
  = \sqrt{2T} 
   \sqrt{1 - \frac{\atd\at}{2T}} \at\\
T^- = &S^- + \dnd\up&
   = 
   \sqrt{2T} \atd 
   \sqrt{1 - \frac{\atd\at}{2T}}
\en
where  $[a,a^{\dagger}] = [\at,\atd] = 1$. 
The relation between $a$, $\up$,$\dn$ and  $\at$, $\f$ and $\p$
then follows from Eq.~(\ref{eqn:newS}) on the magnitude of 
the composite spin, commutation relation for fermionic and bosonic 
operators and the constraints that the total number of particles be 
conserved 
\be 
\fd\f + \pd\p &=& \upd \up  + \dnd \dn
\en
and that double or zero occupancy of the electron orbital correspond 
to the same two sets of states state in either representation
\be
\fd\pd = \upd\dnd \qquad etc.
\en
This set of constraints is sufficient to uniquely define the 
transformation, and after lengthly algebra we find :
\be
\label{eqn:trans}
\at &=& a \left[ 1 + \frac{\dnd \dn - \upd \up }{4S}     
   + \frac{3\upd\up - 5\dnd\dn + 2\upd\up\dnd\dn}{32S^2}\right] 
    + \frac{\upd\dn}{\sqrt{2S}}\left(1 - \frac{1}{4S} \right)
    - \frac{\dnd\up}{\sqrt{2S}}\frac{aa}{4S}\\
\f &=& \up\left[1 - \frac{a^{\dagger}a + \dnd\dn}{4S}
       +\frac{3\dnd\dn + 3a^{\dagger}a - 2\dnd\dn a^{\dagger}a - 
       a^{\dagger}a^{\dagger}aa}{32S^2}\right]
       + \frac{\dn a^{\dagger}}{\sqrt{2S}} \left(1 - \frac{1}{4S} 
\right) \\
\p &=& \dn\left[1 - \frac{a^{\dagger}a + \up\upd}{4S}
       +\frac{3\up\upd + 3a^{\dagger}a - 2\up\upd a^{\dagger}a - 
       a^{\dagger}a^{\dagger}aa}{32S^2}\right]
       - \frac{\up a}{\sqrt{2S}} \left(1 - \frac{1}{4S} \right)
\en
To verify the transformation, we checked that it correctly reproduces
the Kondo coupling term on a single site.  In terms of original 
$a$, $\up$, and $\dn$ operators, the Hamiltonian for a Kondo ``atom'' 
reads :
\be 
\label{eqn:H1}
{\mc H}_1 = - \frac{J S}{2} 
   &&\left(
   (1 - \frac{a^{\dagger}a}{S})[\upd \up  - \dnd \dn]
     + \sqrt{\frac{2}{S}} \sqrt{1 - \frac{a^{\dagger}a}{2S}} a 
     \dnd \up
     + \sqrt{\frac{2}{S}} a^{\dagger} \sqrt{1 - 
\frac{a^{\dagger}a}{2S} }
     \upd \dn
    \right) 
\en
Substituting the transformation, we indeed found that in terms of new
$\at$, $\f$ and $\p$ it exactly reproduces Eq.~(\ref{eqn:tildeH1}). 

For purposes of developing a calculation scheme based on these
coordinates we need the corresponding inverse transformation.
This is found to be identical up to the 
sign of terms at odd order in $1/\sqrt{S}$ :
\be
\label{eqn:inversea}
a &=& \at \left[ 1 + \frac{\pd\p - \fd\f }{4S} 
      + \frac{3\fd \f - 5\pd \p + 2\fd \f \pd \p}{32S^2} \right] 
    - \frac{\fd \p}{\sqrt{2S}}\left(1 - \frac{1}{4S} \right)
    + \frac{\pd \f}{\sqrt{2S}}\frac{\at\at}{4S}\\
\label{eqn:inverseup}
\up &=& \f\left[1 - \frac{\atd\at + \pd \p}{4S}
       +\frac{3 \pd \p + 3 \atd\at 
       - 2\pd \p \atd\at - 
       \atd\atd\at\at}{32S^2}
       \right] - \frac{\p \atd}{\sqrt{2S}} 
   \left(1 - \frac{1}{4S} \right) \\
\label{eqn:inversedn}
\dn &=& \p\left[1 - \frac{\atd\at + \f \fd}{4S}
       +\frac{3 \f\fd  + 3 \atd\at 
       - 2 \f\fd \atd\at - 
       \atd\atd\at\at}{32S^2}
       \right] 
       + \frac{\f \at}{\sqrt{2S}} 
   \left(1 - \frac{1}{4S} \right)
\en
The inverse transformation 
Eqs.~(\ref{eqn:inversea}--\ref{eqn:inversedn}) can be applied
whenever the Hamiltonian Eq.~(\ref{eqn:KondoH}) has a magnetically
ordered ground state, and since all Fermi and Bose operators
are well defined, is an ideal starting point for constructing 
a diagrammatic perturbation theory of spin and charge excitations
in such a system, and for calculating response functions of the 
DE model in a controlled way.  

For the simplest possible case of two core spins sharing 
a single electron, the model Eq.~(\ref{eqn:KondoH}) can
be solved exactly~\cite{anderson}, and one of us has checked 
explicitly that our expansion scheme reproduces all
features of the exact solution, order by order in 
$1/\sqrt{S}$~\cite{nsps}.

\section{spin wave excitations in the Double Exchange model}
\label{DEFM}

We now return to the DE model and analyze the form 
of the spin wave dispersion using the transformation derived above.
As discussed, for a ferromagnetic Kondo coupling $J_H$ and 
$t/J_H \rightarrow 0$, $p-$operators can be safely neglected. 
The effective Hamiltonian for the DEFM 
is therefore found by substituting the inverse transformation 
Eqs.~(\ref{eqn:inversea}--\ref{eqn:inversedn}) 
into Eq.~(\ref{eqn:KondoH}) and dropping 
all $\p$ operators.  Corrections at finite $t/J_H$ will be discussed 
elsewhere.  Substituting the transformation into the DE Hamiltonian 
${\mc H} = {\mc H}_0 + {\mc H}_1$ and expanding in $1/S$ we obtain 
in terms of the new $\f$ and $\at$ operators
\be
{\mc H} &=& {\mc H}_0^{\prime} + {\mc H}_2 + {\mc H}_4
   + {\mc O}(1/S^3)\\
{\mc H}_0^{\prime} &=& -zt \sum_{\langle ij \rangle} \fd_i \f_j\\
{\mc H}_2 &=& \frac{zt}{4S} \sum_{\langle ij \rangle} \fd_i \f_j
    \left[ \left(\atd_i \at_i +  \atd_j \at_j\right)
    \left(1 - \frac{3}{8S} \right)  
    - 2\atd_i \at_j \left(1 - \frac{1}{2S} \right)
    \right]\\
{\mc H}_4 &=& \frac{zt}{32S^2} \sum_{\langle ij \rangle}
    \fd_i \f_j \left[\atd_i \atd_i \at_i \at_i
    + \atd_j \atd_j \at_j \at_j
    - 2 \atd_i \at_i \atd_j \at_j
    \right]
\en
On Fourier transform we obtain :
\be
\label{eqn:effectiveH}
{\mc H}_0^{\prime} &=&  \sum_{k_1} (\epsilon_1 -\mu) \fd_1 \f_1\\
{\mc H}_2 &=& \frac{1}{N} \sum_{k_1\ldots k_4}  v^{13}_{24}
 \fd_1 \f_2 \atd_3 \at_4 \delta_{1+3-2-4} \\
&& v^{13}_{24} =  \frac{1}{4(S+\frac{1}{2})}
   \left[ 
   \left( 1 + \frac{1}{8S} \right) 
   \left( \epsilon_{1+3} + \epsilon_{2+4}\right)
   - \left(\epsilon_1 + \epsilon_2\right)
   \right]\\
{\mc H}_4 &=& \frac{1}{32S^2} \frac{1}{N^2} 
   \sum_{\vec{k}_1\ldots \vec{k}_6} 
   \fd_1 \f_2 \atd_3 \atd_4 
   \at_5 \at_6 . v^{134}_{256}. 
   \delta_{1+3+4-2-5-6}\\
&& v^{134}_{256} = 
   \frac{1}{4} \left[ 
      \epsilon_{1+3-5} + \epsilon_{1+3-6}
      + \epsilon_{1+4-5} + \epsilon_{1+4-6} - 4\epsilon_1
   \right. \no\\ && \qquad \qquad\left.
      \quad \epsilon_{2+6-4} + \epsilon_{2+6-3}
      + \epsilon_{2+5-4} + \epsilon_{2+5-3}
      - 4\epsilon_2
   \right]
\en
All energy scales are set by the electron dispersion,
which for the simple tight binding kinetic energy term 
Eq.~(\ref{eqn:H0}), is given by $\epsilon_k = -zt\gamma_k$.

We see that the original problem of spinfull electrons interacting 
with quantum mechanical localized spins reduces to a single band of 
spinless fermions interacting with a reservoir of (initially dispersionless) 
bosonic spin modes.  For this last problem we can evaluate all quantities 
of physical interest diagrammatically, starting from the bare bosonic 
Green's function given by
\be
{\mc D}_0(q, i\Omega_n) = (i\Omega_n)^{-1}
\en
and the bare $\f$ electron Green's function 
\be
{\mc G}_{0}(k, i\omega_n) = 
   (i\omega_n - \epsilon_k + \mu)^{-1}
\en

We note that since operators $\atd$ and $\at$ describe fluctuations of 
the {\it composite spin}, the pole in the fully renormalized  
bosonic propagator ${\mc D} (q, i\Omega_n)$ coincides with the pole 
in the transverse spin susceptibility, \ie ${\mc D} (q, i\Omega_n)$
describes true spin wave excitations.  Higher-order terms in the 
Holstein-Primakoff expansion of the composite spin $T$ in terms 
of $\atd$ and $\at$ only give rise to the incoherent background in 
the spin susceptibility but do not affect the pole. 
This separation between the pole and the incoherent background only 
makes sense if the damping of spin waves is negligible small. 
We will show that the spin wave damping appears only at $O(1/S^3)$ 
such that to order $O(1/S^2)$ (to which we will perform controlled 
calculations), nonlinear terms in the Holstein-Primakoff transformation 
for $T$ can be neglected~\cite{comm1}.      
\begin{figure}[tb]
\begin{center}
\leavevmode
\epsfysize 7cm
\epsffile{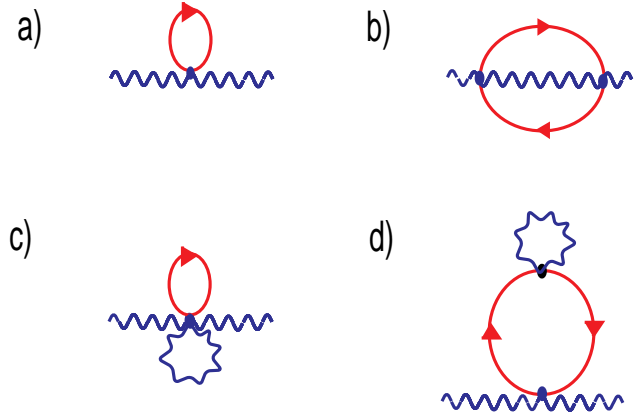}
\caption{a)--d)
Diagrams contributing to spin wave self energy in Ferromagnet to
order ${\mc O}(1/S^2)$.  Only diagrams a) and b) are physically 
relevant. Diagram c) cancels unphysical leftovers from the diagrams 
a) and b),
 and diagram d) accounts for thermal renormalization
 of  the chemical potential and the exchange integral $J_1$ 
but does not change the form of the spin wave dispersion.  }
\label{fig:watermelon}
\end{center}
\end{figure}

\begin{figure}[tb]
\begin{center}
\leavevmode
\epsfysize 7cm
\epsffile{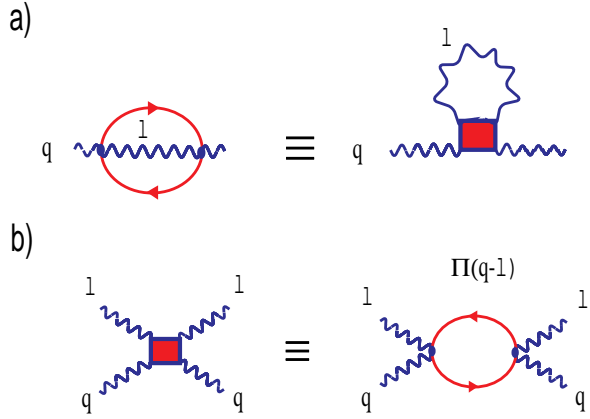}
\caption{
The representation of the diagram Fig~\ref{fig:watermelon}b as
an effective four--boson interaction mediated by the charge 
suscpetibility (particle--hole polarization bubble).}
\label{fig:vertex}
\end{center}
\end{figure}

The dispersionless form of ${\mc D} (q, i\Omega_n)$ does not 
survive self-energy corrections arising from interaction with
Fermions --- the bosonic self-energy depends on momentum $q$, 
and this dependence gives rise to a dispersion of the spin wave 
excitations.  Physically, this dispersion is generated by the 
fact that any departure from perfect ferromagnetic ordering of 
composite spins costs kinetic energy.  
The form of this dispersion should be appropriate to
a ferromagnet on a lattice, \ie it should have a set of Goldstone
modes with energy $\omega_{sw}(q)$ scaling as $q^2$ in the zone 
center, and be continuous across the zone boundary.

The perturbation theory for the
bosonic propagator $D (q, \Omega)$ is straightforward.
We have $D^{-1}(q,\Omega) = \Omega - \Sigma (q, \Omega)$. 
The lowest-order (in $1/S$) contribution to $\Sigma (q, \Omega)$
comes from a  single loop of fermions, and this evaluates to
\be
\label{eqn:cl_disp}
\Sigma^{(1a)} (q) &=& z J_1 S [1-\gamma(\vec{q})]
\en
where
\be
\label{eqn:classical}
J_1 &=& \frac{t}{2S^2} \frac{1}{N}
   \sum_{\vec{k}} n_k
   \gamma(\vec{k})
\en
The result for $J_1$ can be rewritten as 
$J_1 = \overline{t}/(2S^2)$ where
 $\overline{t}$ is the magnitude of the Fermionic 
kinetic energy per bond on the lattice.  

The classical spin wave dispersion of the DEFM, Eq.~\ref{eqn:cl_disp} 
corresponds exactly to what we would expect 
for a nearest neighbour Heisenberg model with exchange integral $J_1$.
The equivalence of the DEFM and the nearest neighbour Heisenberg model
at ${\mc O}(1/S)$ was first discussed by de Gennes \cite{degennes} 
and results for classical spin wave dispersion have been rederived by 
many other authors since \cite{kubo,furukawa,millis,plakida,golosov2}. 

We can translate the energy scale for zero temperature spin wave 
dispersion into a mean field transition temperature for a Heisenberg 
FM using the relation
\be
T_c &=& \frac{zS(S+1)}{3} J_1
\en
giving approximate transition temperature for a $d$ dimensional 
quarter filled cubic lattice with bandwidth 
These estimates are in surprisingly good correspondence with 
transition temperatures for the real CMR materials. 

We now demonstrate  that the DE and Heisenberg models are {\it not} 
equivalent beyond ${\mc O}(1/S)$. To check this, we need to go a step
further and compute the  spin wave dispersion to order ${\mc O}(1/S^2)$. 
For this, we must include the ${\mc O}(1/S^2)$ 
contribution of the one--loop diagram Fig.~\ref{fig:watermelon}a 
and also two--loop self--energy diagrams.  The two--loop diagrams 
are presented in Fig.~\ref{fig:watermelon}b--d.
The first of these diagrams represents the bosonic self--energy due to
the effective four--boson interaction mediated by the Pauli 
susceptibility of fermions.  
The diagram in Fig.~\ref{fig:watermelon}c is the first--order 
contribution from the six--fold, $1/S^2$ term in 
Eq.~(\ref{eqn:effectiveH}).   This diagram is physically 
uninteresting, and is only necessary to restore 
the Goldstone theorem.  The diagram Fig.~\ref{fig:watermelon}d
accounts for thermal renormalization of the chemical potential and the
exchange integral $J_1$ but does not change the functional form of
the spin wave dispersion.  We neglect this diagram below.

We see that from physics perspective, the relevant diagram is 
Fig.~\ref{fig:watermelon}b.  This diagram can be thought of as a 
first order bosonic self--energy due to effective four--boson 
interaction mediated by fermions (see Fig.~\ref{fig:vertex}a--b).
The relation between DEFM and the Heisenberg model then can be 
readily understood on general grounds.  Indeed, in a HFM, 
the four--boson vertex does not depend on frequency and scales as 
$ J_1 q^2 l^2$ (Ref.~\cite{sw}) where $q,l \ll 1$ are the bosonic 
momenta.  In the DEFM, the interaction is mediated by 
the dynamical charge susceptibility of the Fermi gas,
$\Pi (q-l,\omega_p -\omega_l)$ (see Fig.~\ref{fig:vertex}(b)), 
which is generally a complex function of momentum and frequency.
This gives rise to two effects which both are relevant to our 
analysis:

(i)  $\Pi (q-l,\omega_p -\omega_l)$ has a branch cut, which gives 
rise to a nonzero renormalization of the spin wave dispersion at $T=0$, 
and

(ii) the static $\Pi (q-l,0)$, relevant to thermal corrections to the
dispersion at $T < J_1S$,  scales differently from $J_1$ and this changes 
the scale of the self--energy in the DEFM relative to a classically 
equivalent HFM.   In particular, at small $p_F$,  $J_1 \propto p^3_F$, 
while $\Pi (q-l) \propto p_F$ at  $q,l_{typ} \ll p_F$ and 
$\Pi (q-l) \propto p^3_F/|q-l|^2$ for $|q-l| \gg p_F$. 
Thermal corrections to spin wave dispersion in the DEFM 
are therefore enhanced relative to the HFM. 
Furthermore, at small doping $x$,
when typical $q$ and $l$ exceed $p_F$, the momentum dependence of 
$\Pi (q-l)$ causes thermal self energy correction in the DEFM to have
a different functional form from that in the HFM.
     
\begin{figure}[tb]
\begin{center}
\leavevmode
\epsfysize = 75.0mm
\epsffile{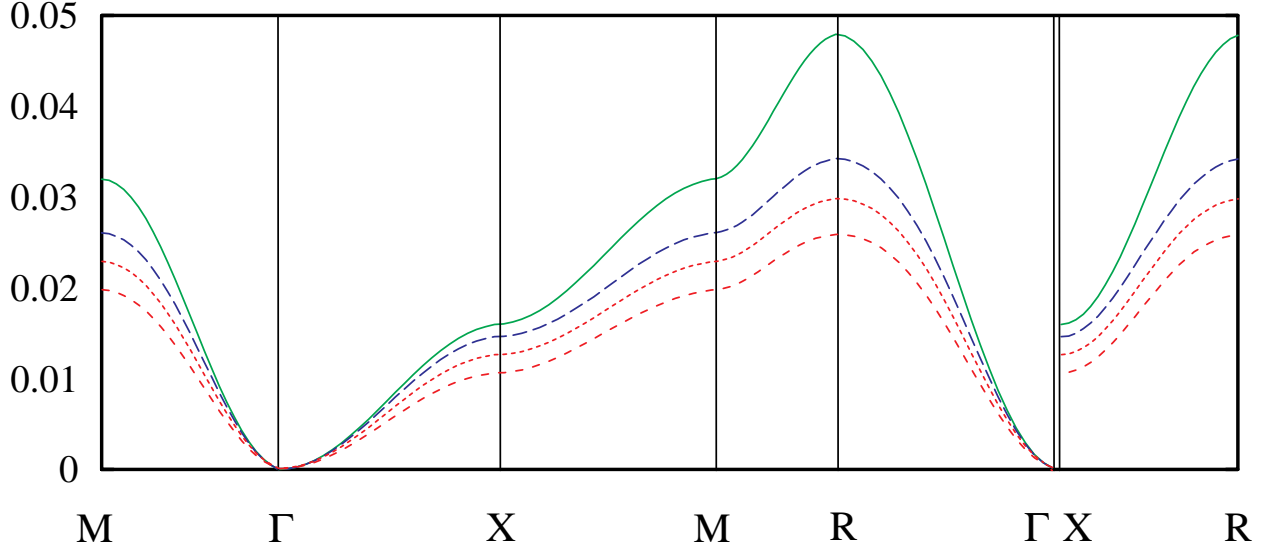}
\caption{
Dispersion of spin waves in a cubic DEFM with electron 
doping $x=0.7$ and spin $S=3/2$, in units of the electron bandwidth 
$2zt=1.0$.
Solid line --- dispersion of classically equivalent Heisenberg model.
Dashed lines --- dispersion of DEFM including leading quantum 
and thermal corrections for (top to bottom) $T=0$, $T=2D/3$ and 
$T = D$ where $D=12J_1S$ is the bandwidth of spin wave excitations.
}
\label{fig:spectrum}
\end{center}
\end{figure}

We now proceed with the calculations.
Assembling all three ${\mc O}(1/S^2)$ contributions to the 
self-energy and splitting it into quantum and thermal pieces,
we obtain after some algebra that all unwanted terms are canceled 
out and the Goldstone mode at $q=0$ survives, as required. 
The resulting  self--energy is given by  
\be
\label{twopieces}
\Sigma^{(2)} (q,\omega) \approx \Sigma^{(2)}(q) 
   &=& \Sigma^{(2)}_{T=0}(q) + \Sigma^{(2)}_T (q)
\en
where
\be
\Sigma^{(2)}_{T=0}(q)
    &=& -\frac{zt}{4S^2} \frac{1}{N^2}
            \sum_{p l} n_F (p) [1 - n_F (l)] 
   \left[(1 -\gamma_q) \gamma_p
        \frac{\gamma_p + \gamma_l}{\gamma_p - \gamma_l} - 
   \frac{\gamma^2_p -
        \gamma^2_{k+p}}{\gamma_p - \gamma_l}
    \right]
\label{T=0}
\en
and
\be
\Sigma^{(2)}_T (q) 
   &=& -\frac{zt}{4S^2} \frac{1}{N^2} n_F (p) n_B (l) 
   \left[\frac{(\gamma_p - \gamma_{p+q})(\gamma_{p+q-l} -
   \gamma_{p+q})}{\gamma_p - \gamma_{p+q-l}} 
   + \frac{(\gamma_p - \gamma_{p+l})(\gamma{p+l-q} -
   \gamma_{p+l})}{\gamma_p - \gamma_{p+l-q}} \right]
\label{T}
\en
Here $n_F (k) = n_F (\epsilon_k)$ and $n_B (q) = n_B (\Omega (q))$
are Fermi and Bose distribution functions, respectively. 
The frequency dependence of the self--energy is relevant for the 
computations of the bosonic damping which is of order $1/S^3$ (see 
below), but can be neglected in the computations to order $1/S^2$ as typical 
bosonic $\omega$ are of order $\omega_{sf} (q) \propto t/S$ and are 
small by $1/S$ compared to typical fermionic energies which are of 
order $t$.  In this situation, the full bosonic dispersion is simply 
\be
\label{dispersion}
\omega_{sw} (q) = z J_1S (1-\gamma_q) + \Sigma^2 (q).
\en 
We will also neglect the temperature dependence of the Fermi functions 
as we are interested in temperatures of order the spin wave bandwidth 
$T \sim t/S \ll t$.
This makes possible the simple separation of the self energy into
thermal and quantum pieces as written in Eq.~(\ref{twopieces}).
We now analyze quantum and thermal corrections to spin wave 
dispersion.  We begin with the case of zero temperature.

\subsection{Quantum corrections at $T=0$}

\subsubsection{spin wave dispersion}

We see from (\ref{T=0}) that at $T=0$, the interaction with the 
charge degrees of freedom leads to an overall reduction in the 
spin wave bandwidth (the first term in (\ref{T=0})), together
with a modification in the form of the dispersion provided by the 
second term in in (\ref{T=0}).  This second term 
\be
\Delta\Sigma^{(2)}(q)  &=& \frac{zt}{4S^2} \frac{1}{N^2}
            \sum_{p l} n_F (p) [1 - n_F (l)] 
    \frac{\gamma^2_p -
        \gamma^2_{k+p}}{\gamma_p - \gamma_l}
\label{T=00}
\en
is either positive or negative throughout the Brillouin zone, 
depending on doping, and has the symmetry 
$\Delta\Sigma^{(2)}(q) = \Delta\Sigma^{(2)}(\pi - q)$ which bare 
spin wave dispersion does not possess.  Near the center of the 
Brillouin zone it behaves as
\be 
\label{tt}
\Delta\Sigma^{(2)}(q\approx 0) \sim \frac{t}{2S^2}I(x)q^2
\en
where 
\be
\label{ttt}
I(x) = \frac{1}{N^2} \sum_{k_1k_2}
   \frac{\gamma_1^2}{\gamma_1 - \gamma_2}
   \left[1-n_F(k_2)\right]n_F(k_1) 
\en
is a doping dependent constant.   By symmetry, the behaviour 
of $\Delta\Sigma^{(2)}(q)$ near the zone corner $q\approx \pi$ 
must have exactly the same form.  Along the zone diagonal 
\be
\Delta\Sigma^{(2)}(q) 
   &=& \frac{zt}{4S^2}I(x) \left[ 1 - \cos(2q) \right]
\en
This form is very similar to the form of correction to Heisenberg 
dispersion found in numerical studies of the DE model on a 
ring \cite{kaplan}.

For a small density of electrons ($x<0.31$), $I(x) > 0$ in both 2D 
and 3D, and corrections to dispersion lead to a ``softening'' of modes
at the zone boundary relative to those the zone center.  The same is 
true for small density of holes $1-x < 0.07$ in $2D$ and 
$1-x < 0.06$ in $3D$.  For intermediate densities
$0.31(7) < x < 0.92(7)$ in $2D$, and $0.31(7) < x < 0.94(2)$ in $3D$,
$I(x)$ is negative, \ie quantum effects cause a relative softening 
of spin wave modes at $q=0$.  These results are in good agreement with 
earlier numerical studies studies~\cite{t=0}.
We discuss the consequences of the non--monotonic doping dependence 
of $I(x)$ in more detail later in Sec.~\ref{orderfrom} when we consider 
order from disorder effects.

We note in passing that if we formally extend our $1/S$
results to arbitrary $S$, the spin wave dispersion becomes unstable 
for $S < S_{cr}$ where in 3D 
$S_{cr} = (1/2N) \sum_k (1+\gamma_k)/(1-\gamma_k) \approx 1$. 
This opens up a possibility that for small S (e.g., $S=1/2$), 
the ground state may not be a ferromagnet, as suggested by some 
numerical studies~\cite{zang}.   However, the extension to small 
$S$ is beyond the scope of the present paper.  

\begin{figure}[tb]
\begin{center}
\leavevmode
\epsfysize = 75.0mm
\epsffile{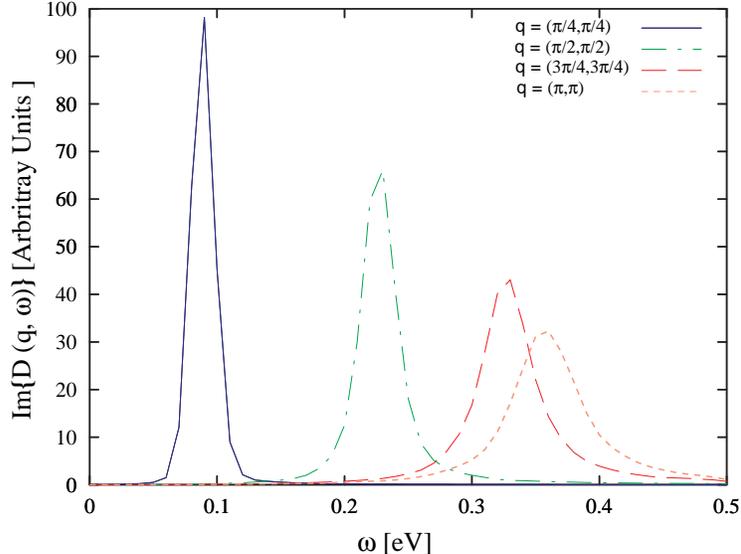}
\caption{
Numerical results for the imaginary part of $D(q,\omega)$
of a $2D$ double exchange ferromagnet with $S=3/2$ and doping 
of $0.18$ holes per site, with energy in units of the hopping 
integral $t$.  The plots show the increasing damping of 
spin waves along the zone diagonal.  Sharp structure in the data
is an artifact of finite numerical resolution and has been partially
removed by convolution with a Gaussian of width $0.01t$.
}
\label{fig:neutrons}
\end{center}
\end{figure}

\subsubsection{Damping of spin waves}

While spin waves are {\it exact eigenstates} of the Heisenberg 
Hamiltonian~(\ref{eq:HFM}), in the DEFM only the $q=0$  
(Goldstone) mode is an eigenstate.
All other spin wave modes in the DEFM  have a finite 
lifetime, even at zero temperature.
Physically, the possibility for a spin wave to decay at $T=0$
is again related to the presence of Fermions.  
A spin wave with energy $\omega_{sw}(q)$ can give energy to a 
particle--hole pair, and decay into a lower energy spin wave
--- the process described by the diagram Fig.~\ref{fig:watermelon}b.
As a consequence $\Sigma^{(2)} (q,\Omega)$ contains an imaginary part 
which describes the damping of a spin wave.

To calculate damping effects it is necessary to keep
the bosonic energies $\Omega$, $\omega_q^{'}$ together with 
the fermionic $\epsilon_k$ in the denominator of $\Sigma^{(2)} (q,\Omega)$.  
This means that the calculations are formally less controlled, as we 
include terms one order smaller in $1/S$.  To work
rigorously at this order it would be necessary to extend the
(inverse) transformation Eq.(\ref{eqn:inversea}--\ref{eqn:inversedn}) 
to ${\mc O}(1/S^3)$ --- a very involved task.
However the physical origin of the damping is clear,
and we verified that the imaginary part of 
the self--energy to ${\mc 0}(1/S^3)$ is fully captured by just 
keeping $\Omega$ in denominators of two--loop self--energy 
diagrams.  We then obtain
\be
\label{eqn:freq}
\Sigma^{\prime\prime}_{T=0}(q,\Omega) 
   &=& \Theta(\Omega) \frac{\pi}{(4S)^2}
       \frac{1}{N^2} \sum_{kq'} 
       \left(\epsilon_{k+q'} + \epsilon_k - 2\epsilon_{k+q}\right)^2
       n_F(k+q')[1-n_F(k)]
       \delta\left((\Omega - \omega_{q-q'}) 
              - (\epsilon_{k+q'} - \epsilon_{k})\right)
\en
where we have grouped together terms of ${\mc O}(0)$
and terms of ${\mc O}(1/S)$ within the argument of 
the delta function.

Without $\Sigma^{\prime\prime} (q,\Omega)$, the  imaginary part of 
the bosonic propagator $D(q,\Omega)$ is just 
a delta function dispersing with the spin wave energy 
$\omega_{sw}(q)$, \ie
\be 
\label{eq:ImD}
D^{\prime \prime} (q,\Omega) &=& -\pi \delta(\Omega - \omega_{sw}(q))
\en
However because of the non--zero $\Sigma^{\prime\prime} (q,\Omega)$, 
the simple delta function peak~(\ref{eq:ImD}) is replaced by a broader 
dispersing feature with is approximately Lorentzian in shape with the 
maximum at $\omega_{sw}(q)$.

We can obtain an estimate of the width of $D^{\prime \prime} (q,\Omega)$ 
by calculating the damping of spin waves on the mass shell 
\be
\label{eqn:gamma}
\Gamma_q &=& \frac{\pi}{4S^2}
       \frac{1}{N^2} \sum_{kq'} 
       \left(\epsilon_k - \epsilon_{k+q}\right)^2
       n_F(k+q')[1-n_F(k)]
       \delta\left((\omega_{sw} (q) - \omega_{sw} (q-q')) 
              - (\epsilon_{k+q'} - \epsilon_{k})\right).
\en
where we have used the delta function on energy to eliminate 
terms of subleading order in $1/S$ from the vertex.
The damping of spin waves must vanish for $q \to 0$, in 
order for the Goldstone theorum to be satisfied. 
For $ q \ll p_F$ we find that $\Gamma_q$ vanishes
as $\omega_{sw} (q) ~q^{d+1} \sim q^{d+3}$.  
A factor $\omega_{sw} (q)$ comes from the 2D integration over 
$d |k-k_F|$ and $d |q^\prime|$, another factor $q^2$ comes from the 
vertex, and the remaining factor $q^{d-1}$ comes from the fact that 
$|q^{\prime}|\leq q$  and hence $d^d q^\prime \sim q^{d-1} d|q^\prime|$.
In 2D, this yields $\Gamma_q \propto q^5$ and in 3D, $\Gamma_q \propto q^6$. 
This fully agrees with earlier estimates~\cite{irkhin,golosov2}. 
This strong dependence on $q$ survives up to $q \sim p_F$. 
At larger $q$, and $p_F \ll 1$, the damping of spin waves  saturates at 
a constant value $\Gamma_{q} \propto (t/S^3)$.  At $p_F \sim 1$,  
$\Gamma_q$ at $q \geq p_F$ is in general a rather complicated function 
of momentum.

In Fig.~\ref{fig:neutrons} we plot the imaginary part of
$D(q,\Omega)$ obtained by numerically evaluating (\ref{eqn:freq}) 
for a set of momenta on the zone diagonal for a DEFM on a 2D square 
lattice, for $S=3/2$ and electron density $x=0.82$.  
From the plot, we see that zone corner modes are clearly 
broadened relative to those in the zone center.  
Evaluating $\Gamma_q$ throughout the BZ for a range of 
dopings in 2D we obtain a functional form of damping similar to 
that recently reported in~\cite{golosov2}.  However our estimate on 
the upper bound for values of damping is about $25\%$ smaller than 
that given in~\cite{golosov2}.

\begin{figure}[tb]
\begin{center}
\leavevmode
\epsfysize = 75.0mm
\epsffile{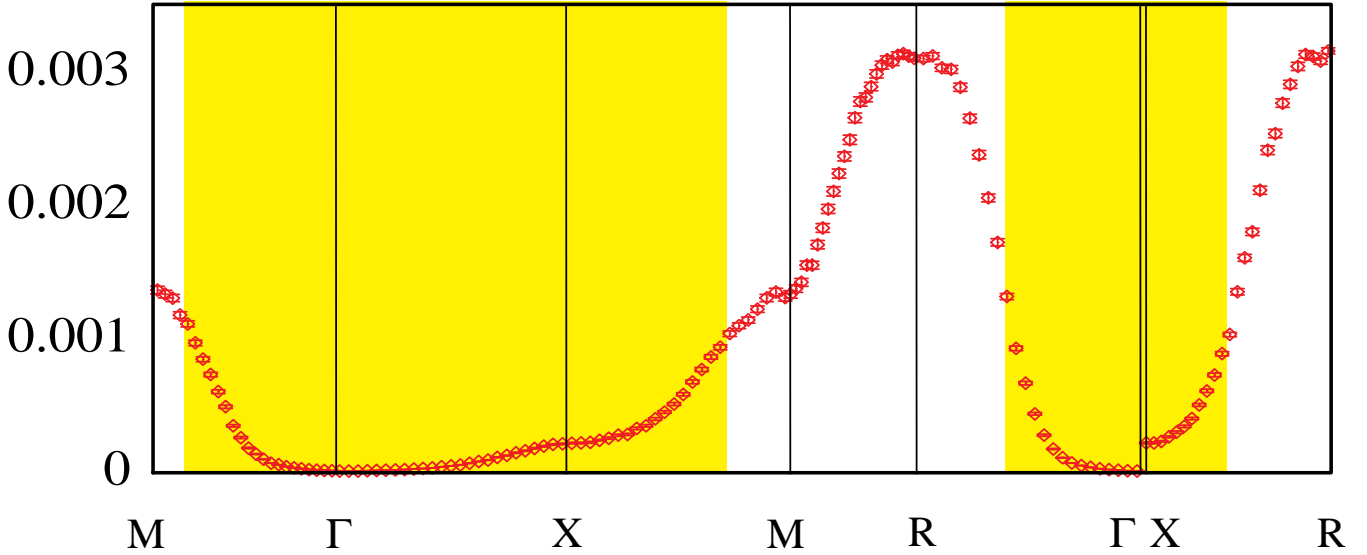}
\caption{
Damping of the spin wave on the mass shell as a function of 
momentum throughout the BZ for a cubic DEFM with  electron 
doping $x=0.7$ and spin $S=3/2$, in units of the electron 
bandwidth $2zt=1.0$.  The volume enclosed by the Fermi surface 
is denoted by shading. 
}
\label{fig:damping}
\end{center}
\end{figure}

The results for the 3D cubic case are similar, and best summarized
by plotting the $\Gamma_q$ along all major symmetry directions.
In Fig.~\ref{fig:damping} we plot the damping for a cubic system 
with electron doping $x=0.7$ and spin $S=3/2$, in units of the 
electron bandwidth $2zt=1.0$.  The damping of spin waves is small 
at low $q$, but strongly momentum dependent.  Damping is large for 
$q$ approaching the zone edge, and has a maximum value of about 
$10 \%$ of the spin wave dispersion at the zone corner.  
Stationary points of the damping (maxima, minima and points of 
inflection) occur at symmetry points of the the BZ or where $q$ 
crosses the Fermi surface. 

The large absolute value of the damping at large $q$ in the DEFM 
and its strong momentum dependence is consistent with the experimental 
behavior seen in Neutron scattering experiments on the Manganites, 
where damping is large, highly momentum dependent, and can rise to 
$\sim 10\%$ of spin wave dispersion at the zone corner~\cite{neutrons}.

\subsection{Thermal corrections at $T>0$}

We now discuss the form of the correction to the spin wave dispersion 
at finite $T$.  Continuing the spirit of comparison with the low $T$ 
expansion for a Heisenberg FM, we assume that $T \leq J_1S \ll t$. 
In a nearest--neighbor Heisenberg FM the thermal renormalization 
of the spin wave dispersion to first order in $1/S$ is 
\be
\Sigma^{(2)}_T (q) 
   &=& - \frac{1}{4S}~\omega_{sw} (q) 
       B\left(\frac{T}{J_1S}\right) 
\en
where~\cite{sw}
\be
B\left(\frac{T}{J_1S}\right) 
   = \frac{1}{N} \sum_l n_B (l) (1 - \gamma_l)
\en
At the lowest temperatures, 
\be
B\left(\frac{T}{J_1S}\right) 
   \approx \frac{\zeta(5/2)}{32 
   \pi^{3/2}}~\left(\frac{T}{J_1S}\right)^{5/2}
   \label{sw1}
\en
We see that the dispersion preserves its $T=0$ form, and that 
thermal fluctuations only reduce the overall scale of dispersion by 
an amount which at low $T$ is proportional to $T^{5/2}$.  For the 
DEFM we find below that at small $x$, the form of $\Sigma^{(2)}_T (q)$ 
is totally different from that in HFM.  For finite $x$, we find 
that $\Sigma^{(2)}_T (q)$  can be crudely approximated by the same 
functional formulas (\ref{sw1}), but the overall factor has completely 
different doping dependence from the effective exchange coupling $J_1$. 

\begin{figure}[tb]
\begin{center}
\leavevmode
\epsfysize = 75.0mm
\epsffile{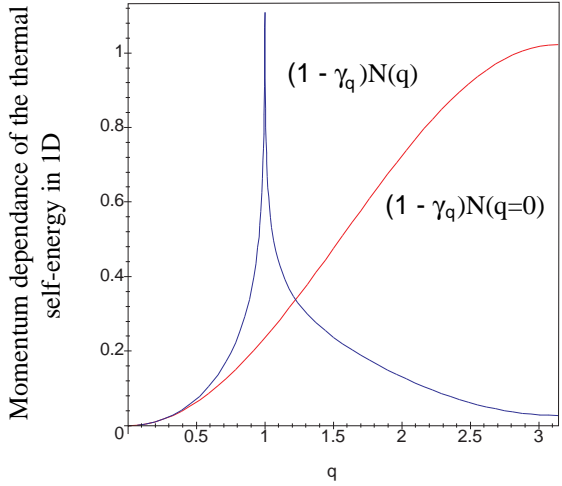}
\caption{
The momentum dependence of the 
thermal correction to spin-wave 
dispersion in a DEFM in 1D for
$p_F = 0.4$, following Eq.~(\ref{new2}).    
}
\label{fig:1D}
\end{center}
\end{figure}

We now proceed with the analysis of Eq.~(\ref{T}). 
In general, at finite $T$ there are three typical momenta in the
problem, the spin wave momentum $q$, the Fermi momentum $p_F$, 
and a thermal length scale $l_{typ} \sim (T/J_1S)^{1/2}$ 
(in units where the lattice constant $a=1$).  
Let us consider first the case when $T$ is very low and $l_{typ} \sim 
(T/J_1S)^{1/2}$ are smaller than  $p_F$. 
This case mimics the behavior at intermediate electron densities. 
Expanding in (\ref{T}) in $l$ we obtain after a simple algebra:
\be
\Sigma^{(2)}_T (q)  = \frac{t}{4 S^2} \sum_l n_B (l) l^2 \sum_p
 n_F (p) \gamma_p \Psi (p,q)
\label{t11}
\en 
where
\be
 \Psi (p,q) = 1 + \gamma_q - 
\frac{4}{z^2} \sum_i 
\frac{\sin^2(p_i + q_i) + \sin^2 (p_i)}{\gamma_p (\gamma_p - 
\gamma_{p+q})}
+ \frac{\sin^2(p_i - q_i) + \sin^2 (p_i)}{\gamma_p (\gamma_p - 
\gamma_{p-q})}
\label{new1}
\en 
where $i$ takes the values $x, y$ and $z$.
If $\Psi (p,q)$ was just proportional to $1- \gamma_q$, 
the thermal self-energy would have the same functional form as in HFM.
We see, however, that $\Psi (p,q)$ is more complex and depends on 
both variables.

The form of $\Psi (p,q)$ is particularly simple in 1D. 
Here we obtain from Eq.~(\ref{new1}) 
\begin{eqnarray}
 \Psi (p,q) &=& (1 + \cos q) - \frac{\sin^2(p + q) + \sin^2 (p)}{\cos 
p 
(\cos p - \cos (p+q))} -\frac{\sin^2(p - q) + \sin^2 (p)}{\cos p 
(\cos 
p - \cos (p-q))} \nonumber \\
&& = - (1 -\cos q) \frac{1 + 2 \sin^2 p + \cos q}{1 - \cos q - 2 
\sin^2 
p}
\end{eqnarray}
We see that at small $q$ and {\it finite} $p$,  
$\Psi (p,q)$  scales as $q^2$ as it indeed should. 
Substituting the result for $\Psi (p,q)$ into (\ref{t11}) we find 
after integrating over $p$
\be
\Sigma^{(2)}_T (q)  = - \frac{t}{4 S^2} \left(\sum_l n_B (l) 
l^2\right)~
 (1 - \cos q)~N(q)
\label{new2}
\en 
where
\be
N(q) = \frac{\sin p_F}{\pi} \left(-1 + \frac{1}{2 \sin p_F \sin 
q/2} 
\log \frac{\sin p_F + \sin q/2}{|\sin p_F - \sin q/2|}\right)
\label{new3}
\en
At $q \rightarrow 0$, 
\be
N(q) \rightarrow \frac{1}{\pi}~\left(\frac{1}{\sin p_F} - \sin p_F 
\right).
\en
At the zone boundary, $q = \pi$,
\be
N(\pi) = \frac{\sin p_F}{\pi} \left(-1 + \frac{1}{2\sin p_F} 
\log{\frac{1+ \sin p_F}{1-\sin p_F}}\right).
\en
At small $p_F$, $N(\pi) = (1/3\pi) \sin^3 p_F + ...$.
We see that $N(q) >0$, and hence $\Sigma^{(2)}_T (q)$ 
is negative, \ie thermal fluctuations reduce the spin wave energy. 
This agrees with the spin wave result for the HFM. 

In Fig.~\ref{fig:1D} we plot the form of thermal corrections to 
spin wave dispersion for a DEFM in 1D with $p_F = 0.4$, together with a 
correction proportional to $1 - \gamma_q$ having the same prefactor at 
small $q$.   
We clearly see two effects --- the suppression of temperature corrections
at large $q$ relative to small $q$, and a logarithmic singularity coming from
the perfect nesting of the Fermi surface in 1D.

The $q^2$ behavior at the smallest $q$  exists in all dimensions 
as we have explicitly verified.  The explicit forms are, however, 
rather complex and we refrain from presenting them.  
It is essential that for all $D$, the prefactor scales as $1/p^2$ 
for $q \ll p \ll 1$.   In 3D, the $q^2/p^2$ form of 
 $\Psi (p,q)$ yields for $q, l \ll p_F \leq 1$,
\be
\Sigma^{(2)}_T (q) = - \frac{t q^2 p_F}{144 \pi^2 S^2} \frac{1}{N}
\sum_l n_B (l) l^2 \propto q^2 p_F T^{5/2}
\label{t111}
\en 
In 2D, we obtained 
\be
\Sigma^{(2)}_T (q) = - \frac{t q^2}{48\pi S^2} \frac{1}{N}
\sum_l n_B (l) l^2 \propto q^2 T^{2}
\label{t111_2}
\en 
Since the overall scale of spin wave dispersion in the DEFM at low 
doping is proportional to $x$, (which scales as $p_F^2$ in 2D and 
$p_F^3$ in 3D), we see that the overall scale of thermal corrections
in the DEFM is {\it enhanced} by a factor $1/p_F^2$ relative to a 
Heisenberg model with the effective FM coupling $J_1$.  

Numerical results for the spin wave dispersion in a cubic system 
with an electron doping of $x=0.7$, including thermal corrections, 
are shown in Fig.~\ref{fig:spectrum}.  Thermal corrections are 
generally small compared with the quantum corrections at $T=0$, 
the two effects becoming comparable only for $T \sim 6J_1S$. 

\begin{figure}[tb]
\begin{center}
\leavevmode
\epsfysize = 75.0mm
\epsffile{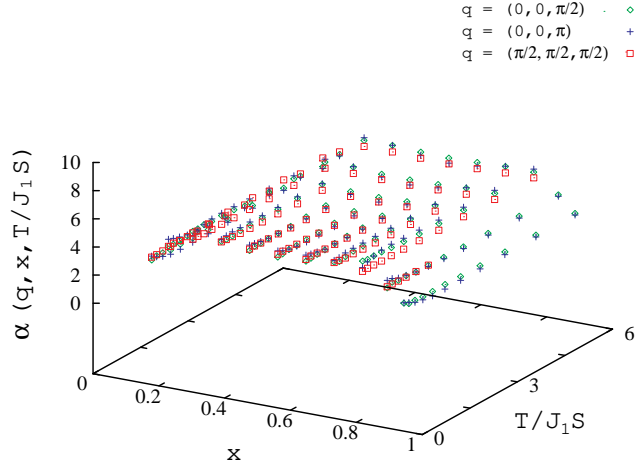}
\caption{
Ratio of thermal corrections to spin wave dispersion in a DEFM
to the thermal corrections of a classically equivalent Heisenberg FM
$\alpha (q,x,T) = \Sigma^{DEFM}_T (q)/\Sigma^{HFM}_T (q)$
for momenta $q=(0,0,\pi/2)$,  $q=(0,0,\pi)$ and 
$q=(\pi/2,\pi/2,\pi/2)$
as a function of doping $x$ and reduced temperature $T/J_1S$.
Plot shows that data for a wide  range of dopings, momenta and 
reduced temperatures approximately collapse  onto a ``universal''
surface.
}
\label{fig:alpha}
\end{center}
\end{figure}

It is helpful to divide thermal self energy corrections in the 
DEFM by the thermal corrections for a Heisenberg model
with the same spin stiffness $J_1S$, \ie to consider 
\be 
\alpha(q,x,T) &=& \frac{4S}{B(T/J_1S)}~
\frac{\Sigma^{(2)}_T (q)}{\omega_{sw}(q)}
\en
We plotted this ratio against reduced temperature $T/J_1S$ for 
several $q$ and and a range of dopings $0.1 < x < 0.9$ in  
Fig.~\ref{fig:alpha}.   
We see that for a wide range of dopings and reduced temperatures 
the data for different $q$ collapse onto a single ``universal'' surface.  
This means that $\alpha$ remains roughly constant as a function of $T/J_1S$, 
and also depends only weakly on on $q$, which implies that for
the experimentally relevant intermediate dopings, the 
thermal self-energy in a DEFM roughly mimics that in a HFM. 
Still, however, the overall amplitudes of the corrections are very 
different for the reasons given above. 
We found that the ratio of the two is approximately constant at
$\alpha \approx 5$ for $0.25 < x < 0.75$, falling 
away to about $\alpha = 2$--$3$ for $x=0.1$ ($x=0.9$).
These numbers should be compared to enhancement of corrections
by a factor $2$--$3$ relative to a nearest neighbour HFM 
observed in La$_{0.85}$Pb$_{0.15}$MnO$_3$~\cite{lynn}, for which 
$x=0.85$.  The authors of~\cite{lynn} interpreted the 
enhancement of temperature corrections in La$_{0.85}$Pb$_{0.15}$MnO$_3$ 
in terms of effective non--nearest neighbour couplings --- we note
that these are dynamically generated by the form of interaction
between spin waves in the DEFM --- see Fig.~\ref{fig:vertex}(b).
At the smallest dopings $\alpha$ acquires a strong dependence 
on both $q$ and $T/J_1S$, as discussed below.
We also found that for $D =2$ and $D=1$, the deviations from 
$\alpha \approx const$ are more prominent for the same intermediate 
$x$.   In particular, in 1D a simple experimentation with trigonometry 
yields $N (\pi) < N(0)$.  This implies that compared to the HFM, 
the strength of thermal fluctuations in DEFM is reduced 
near $q = \pi$.   

We now consider analytically what happens if we lift the restrictions 
that both $q$ and $l_{typ} \sim T/J_1S$ are smaller  than $p_F$.  
As we still focus on low $T \leq J_1 S$, $p_F$ 
should also be small which in turn implies that electron density $x \ll 1$. 
We verified that the results similar to the ones below also hold for small 
hole density $1-x \ll 1$. 

Let us first consider what happens when $q$ exceeds $p_F$. 
In this limit, we  can approximate $\Psi(p,q)$ by $\Psi(0,q)$. 
Substituting the result into the expression for the self-energy, 
we obtain, in any D
\be
\Sigma^{(2)}_T (q)  =  - \frac{t~ x}{4 S^2}
\frac{1}{N}
\sum_l n_B (l) l^2 P(q)
\label{t2}
\en
where 
\be
P(q) = \left[\frac{2}{z}
\frac{1-\gamma_{2q}}{1-\gamma_q} -(1+\gamma_q)\right]
\label{new5}
\en
In 1D, (\ref{t2}) reduces to 
\be
\Sigma^{(2)}_T (q)  =  -\frac{t~ x}{4 S^2}
\frac{1}{N}
\sum_l n_B (l) l^2 ( 1 + \cos q)
\label{new4}
\en
This result coincides with Eq.~(\ref{new2}) as for $p_F \ll 1$ and
$q \gg p_F$, $N(q)$ from (\ref{new3}) reduces to $N(q) = x (1 + \cos 
q)/(1 - cos q)$ where $x = p_F/\pi$.

In 3D, Eqn (\ref{t2}) yields  
$\Sigma^{(2)}_T (q) \propto p^3_F T^{5/2}$. In 2D, we have
$\Sigma^{(2)}_T (q) \propto p^2_F T^{2}$.
Comparing these results with Eq.~(\ref{t111}) and (\ref{t111_2}),
we see that the at $q \gg p_F$, the self-energy is {\it reduced} by 
$p^2_F/q^2$, relative to corrections at the zone center.
This is precisely the result we anticipated on general grounds as 
when $q$ exceeds $p_F$, the effective boson--boson vertex is reduced 
due to the reduction of the charge susceptibility which mediates 
boson--boson interaction. 
Note that this reduction eliminates the quadratic dependence on $q$, 
\ie the thermal self-energy becomes flat at $q \gg p_F$.

Curiously enough, in 3D, at these low dopings, $P(q)$ is negative 
for all $q$ along zone diagonal ($P(q) = -(1/3) (1 - \cos q)$), 
\ie the thermal self-energy changes sign compared to Eq.~(\ref{t111}).
In 2D, $P(q)$ vanishes along zone diagonal, and in 1D, $P(q) = 1 + 
\cos q$ is positive shown above (see Eq.~(\ref{new4}).
\begin{figure}[tb]
\begin{center}
\leavevmode
\epsfysize = 75.0mm
\epsffile{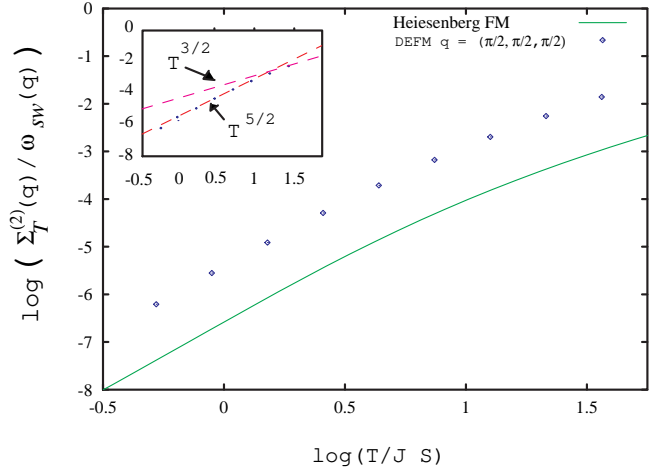}
\caption{
Thermal corrections to spin wave dispersion in a cubic DEFM
with $x=0.67$ for $q=(\pi/2,\pi/2,\pi/2)$ (points) compared with 
those for a classically equivalent Heisenberg FM (solid line).
While the functional form of the corrections is similar, their magnitudes 
are very different.  
The inset  shows the crossover from $T^{5/2}$ behavior at low $T$ 
to $T^{3/2}$ behavior at $T \geq J_1S$.}
\label{fig:finiteT1}
\end{center}
\end{figure}

\begin{figure}[tb]
\begin{center}
\leavevmode
\epsfysize = 75.0mm
\epsffile{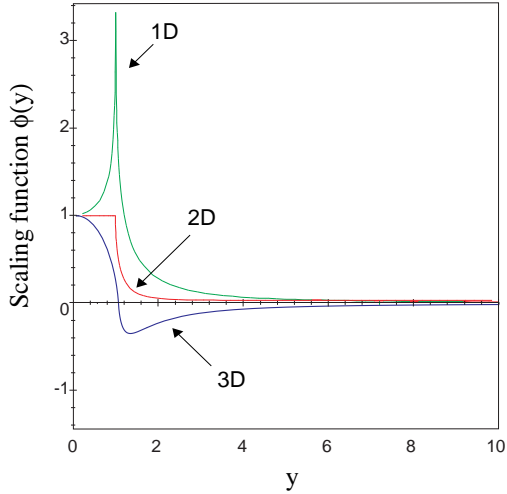}
\caption{
Scaling function governing thermal corrections to spin wave
dispersion in the DEFM in 1D --- Eq.~(\ref{eqn:1D}), 
2D --- Eq.~(\ref{eqn:2D}) and 3D --- Eq.~(\ref{int}).
}
\label{fig:phi3D}
\end{center}
\end{figure}

The reduction of the bosonic self--energy due to the reduction of the 
charge susceptibility at large momenta can also be detected by analyzing 
the form of  $\Sigma^{(2)}_T$ when $q$ is  smaller than $p_F$,
but  $l_{typ}$ exceeds $p_F$. 	
This case is even more instructive than the large $q$ and small 
$l_{typ}$ limit, as $\Sigma^{(2)}_T$ can be obtained analytically for 
{\it arbitrary} ratio of $l_{typ}$ and $p_F$. 
Indeed, expanding in all three momenta in (\ref{T}) we obtained in 3D
\be
\Sigma^{(2)}_T (q)  = - \frac{t q^2 p_F}{144 \pi^2 S^2} \frac{1}{N}
\sum_l n_B (l) l^2 \Phi_{3D} \left(\frac{l}{2\p_F}\right)
\label{t1}
\en
The scaling function  $\Phi_{3D}(y)$ is given by 
\be
\Phi_{3D} (y) = \frac{3}{y^2} \left[ \int_{z_{min}}^1 
dz \sqrt{1 - y^2(1-z^2)} (1+2 y^2 z^2) -1\right]
\label{int}
\en
For $y<1$, $z_{min} =0$, for $y>1$,  $z_{min} = \sqrt{(y^2-1)/y^2}$. 
Evaluating the integral analytically in the limiting cases, we find  
$\Phi_{3D} (y \rightarrow 0) =1,~\Phi_{3D} (y=1) =0$, 
$\Phi_{3D} (y \gg 1) \approx -1/y^2$. 
The $y=0$ limit reproduces Eq.~(\ref{t111}). For large $y$, 
\ie for $ q \ll p_F \ll (T/J_1 S)^{1/2}$, we have 
$\Sigma^{(2)}_T (q)  = C q^2 p^3_F (T/J_1S)^{3/2}$ where $C>0$. 
The scaling function $\Phi_{3D} (y)$ is shown in Fig. (\ref{fig:phi3D}).
The sign change of $\Phi_{3D} (y)$ between small and large $y$ 
collaborates our earlier observation that in 3D, the self-energy at 
the smallest $T$ and at $T/J_1S \geq p_F$ have different signs.
When $q$ exceeds $p_F$ and become comparable to $(T/J_1S)^{1/2}$, 
Eq.~(\ref{t1}) becomes invalid, and $\Sigma^{(2)}_T (q)$  smoothly 
interpolates into $\Sigma^{(2)}_T (q) \propto  p^3_F T^{5/2}$, as in 
Eq.~(\ref{t2}).
 
\begin{figure}[tb]
\begin{center}
\leavevmode
\epsfysize = 75.0mm
\epsffile{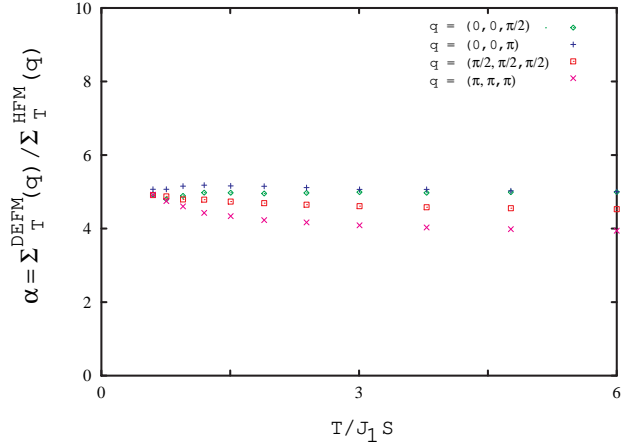}
\caption{
Thermal corrections to spin wave dispersion in a DEFM
with $x=0.67$ normalized to those in an equivalent HFM for 
$q=(0,0,\pi/2)$, $q=(0,0,\pi)$,$q=(\pi/2,\pi/2,\pi/2)$ and 
$q=(\pi,\pi,\pi)$.
}
\label{fig:finiteT2}
\end{center}
\end{figure}

A very similar behavior holds in 2D case.  Here we obtained
\be
\Sigma^{(2)}_T (q)  = - \frac{t q^2}{48 \pi S^2} \frac{1}{N}
\sum_l n_B (l) l^2 \Phi_{2D} \left(\frac{l}{2\p_F}\right)
\label{2t}
\en
where 
\be
\label{eqn:2D}
\Phi_{2D}(y) &=& 1,~~y<1 \nonumber \\
\Phi_{2D}(y) &=& -\frac{3}{2 y^2} + \frac{1}{\pi y^3} 
\int_0^{\phi_{max}} 
\frac{d\phi}{\cos\phi} \sqrt{1 - (y \sin{\phi})^2} (1 +2y^2 + 4 
(y\cos \phi)^2),~~y>1
\label{int2}
\en
where $\sin{\phi_{max}} =1/y$. In the limit $y\gg 1$, we found 
analytically $\Phi_{2D}(y) =a/y^4$ with $a>0$. 

Again, when $q$ exceeds $p_F$ and becomes comparable to : 
$(T/J^*S)^{1/2}$, Eq.~(\ref{2t}) becomes invalid, and 
$\Sigma^{(2)}_T (q)$ smoothly interpolates into 
$\Sigma^{(2)}_T (q) \propto p^3_F T^{5/2}$ as in Eq.(\ref{t2}).

Finally, in 1D,
\be
\Sigma^{(2)}_T (q)  = - \frac{t q^2}{8\pi S^2 p_F} \frac{1}{N}
\sum_l n_B (l) l^2 \Phi_{1D} \left(\frac{l}{2\p_F}\right)
\label{1t}
\en
where 
\be
\label{eqn:1D}
\Phi_{1D} (y) = \frac{1}{2y} \log{\frac{1+y}{|1-y|}}.
\en
In the two limits, $\Phi_{1D} (0) =1$, and 
$\Phi_{1D} (y \gg 1) = 1/y^2$.  As before, when $q$ exceeds 
$p_F$ and becomes comparable to $(T/J^*S)^{1/2}$, Eq.~(\ref{1t}) 
smoothly interpolates into Eq.~(\ref{new4}). 

We see  that in any D, the Heisenberg form of the correction to the 
spin wave dispersion is reproduced only at small 
$(T/J_1S)^{1/2} \ll p_F$, when the scaling function 
$\Phi (y)$ can be approximated by a constant.  
At larger temperatures $y_{typ} \sim (T/J_1S p^2_F)^{1/2}$ 
become large, and the variation of $\Phi(y)$ cannot be neglected.  
The only difference between different $D$ is that 
$\Phi_{3D}(y)$ changes sign at $y =O(1)$ and becomes negative at 
large $y$, while $\Phi_{2D}$ and $\Phi_{1D}$ remain positive for 
all $y$. 

Unfortunately, the new physics we just described is confined 
to very small $x$in D=3. This is simply because in 3D 
$p_F = (6\pi^2 x)^{1/3}$, and to make $p_F$ small, one needs 
a {\it very} small x. 
For such small $x$, our numerical analysis is not very accurate, but 
we were able to confirm that $\Sigma^{(2)}_T (q)$ changes sign as $T$ 
increases.  We also clearly detected the downturn renormalization of 
$|\Sigma^{(2)}_T (q)|$ with increasing $T$ at $x$ (or $1-x$) $\sim 0.1$.  

\section{Order from disorder in the double exchange model}
\label{orderfrom}

We now apply the results of the previous section to study the order 
from disorder phenomenon in the double exchange model.  We recall that 
if double exchange ferromagnetism $J_1$ is put into competition with 
direct antiferromagnetic superexchange $J_2$ between core spins, 
then classically, ferromagnetic state becomes unstable at $J_1 < 
J_2$, but the intermediate configuration is not specified --- the 
only requirement on classical spins is that neighboring spins are 
misaligned by an angle $\Theta$: 
\be
\label{eqn:angle}
\cos \Theta = J_1/J_2.
\en 
This condition allows an infinite set
of intermediate states. The two limits of this set are canted and
spiral spin configurations, as discussed in the introduction.

We first demonstrate that at a classical level, the spin wave 
excitation spectrum of either canted or spiral state indeed contains
a zero-energy branch of excitations. Consider for definiteness a 
canted phase.
The bosonic  Hamiltonian for a two sublattice
canted structure described by the full  ${\mc H}_0 + {\mc H}_1 +
 {\mc H}_2$ at vanishing $t/J_H$ and $J_2/J_H$ can be obtained 
directly within our expansion scheme. 
Working to ${\mc O}(1/S)$ and introducing two
quantization axes misaligned by $\theta/2$, we obtain
\be
{\mc H}_{\theta} 
   &=& {\mc H}^{(0)}_{\theta}  
   + {\mc H}^{(1)}_{\theta}  + {\mc H}^{(2)}_{\theta}  + \ldots
\en
where 
\be
\label{eqn:cant1}
{\mc H}^{(0)}_{\theta} 
   &=& \sum_{\langle ij\rangle} 
   \left\{ 
      -t \cos \frac{\theta}{2} \left[ \fd_{Ai}\f_{Bj} + \hc \right]
      + J_2 S^2 \cos\theta
   \right\} \\
{\mc H}^{(1)}_{\theta} 
   &=& \sum_{\langle ij\rangle} 
   \left\{ 
      \frac{t}{\sqrt{2S}} \sin\frac{\theta}{2} \left[
         \fd_{Ai}\f_{Bj} \left(\tilde{b}_j - \atd_i \right)
         + \hc \right] \right. \no\\
   &&  - J_2S \sqrt{\frac{S}{2}} \left. \sin \theta 
         \left[
             \left( \tilde{b}_j - \atd_i \right)
         + \hc 
         \right]
   \right\} \\
{\mc H}^{(2)}_{\theta} 
   &=&  \sum_{\langle ij\rangle} 
   \left\{ 
      \frac{t}{4S} \cos\frac{\theta}{2} \left[
         \fd_{Ai}\f_{Bj} \left(
         \atd_i\at_i + \tilde{b}^{\dagger}_j \tilde{b}_j 
         - 2 \atd_i \tilde{b}_j
         \right)
         + \hc \right] \right. \no\\
   &&  + J_2S \left. \left[
         \cos\theta \left(
            \atd_i \at + \tilde{b}^{\dagger}_j \tilde{b}_j
            \right)
         - \cos^2 \frac{\theta}{2} \left(
            \atd_i \tilde{b}_j + \tilde{b}^{\dagger}_j \at_i
            \right)
         + \sin^2 \frac{\theta}{2} \left(
            \at_i \tilde{b}_j + \tilde{b}^{\dagger}_j \atd_i
            \right)
         \right]
   \right\}  
\en
where $\f_A$ denotes a local ``up spin'' electron on sublattice
$A$, $\tilde{b}$ a Holstein-Primakoff boson on sublattice $B$, etc.,
and the sum over $\langle ij \rangle$ is restricted to pairs 
of neighbouring sites with site $i$ on the $A$ sublattice and 
site $j$ on the $B$ sublattice.   We note that the effective 
Hamiltonian for the antiferromagnetic
N{\'e}el state is obtained by setting $\theta = \pi$ above.

To zeroth order in $1/S$ the ground state energy of this 
configuration is 
\be
E_0 &=& 
   2Nz \left[J_2S^2 \cos\theta 
            - \overline{t} \cos\frac{\theta}{2}
       \right]
\en
Minimizing this with respect to $\theta$ gives the de Gennes result
(Eq.~\ref{eqn:angle}) for the canting angle. 

The spin wave dispersion to leading order in $1/S$ is obtained
in the same way as for a ferromagnet, by evaluating the leading 
corrections to an (initially dispersionless) bosonic propagator. 
This spin wave ``spectrum'' is doubly degenerate,
and the lattice momentum $\vec{q}$ is restricted to the reduced 
Brillouin appropriate to the two sublattice spin configuration.
The evaluation of the self-energy to order $1/S$ is
technically a bit more involved than for a ferromagnet as in addition 
to the one-loop self-energy from ${\mc H}^{(2)}_{\theta}$ we also need to
compute the second-order contribution from ${\mc H}^{(1)}_{\theta}$. 
The last term vanishes at zero momentum after one substitutes the 
canting angle, as it indeed should, but it does contribute at 
second order. We found that the two contributions exactly cancel each 
other, \ie to order $1/S$, 
\be
\omega_{sw}(\vec{q}) &\equiv& 0
\en
The vanishing of the leading order spin wave dispersion is clearly
linked to the degeneracy of the classical groundstate, and, as we 
already mentioned in the introduction, has a simple 
physical interpretation --- at the classical level we may take any
spin on the $B$ sublattice and rotate it about the direction of 
magnetization of the $A$ sublattice without altering the angle 
between it and the neighbouring spins.  As $\theta$ does not change, 
this perturbation does not cost any energy, and by extension the 
spin wave dispersion at ${\mc O} (1/S)$ is zero~\cite{degennes,golosov1}.

We now proceed to the computations beyond the leading order in $1/S$.
Our key goal is to understand which configuration of the classically 
degenerate set is selected by either quantum or thermal fluctuations. 
There are two ways to attack this problem. First, we may continue 
with the analysis of the
canted phase and perform computations to next order in $1/S$. This is
possible but requires a lot of efforts as at each order 
there is a competition between the contributions from from the term 
quartic and bosons and the term linear in bosons. 
A second route which we choose is to analyze the momentum of the 
instability of the ferromagnetic configuration.  
To ${\mc O}(1/S)$, the excitation 
spectrum of a ferromagnet in the presence of the antiferromagnetic 
exchange is given by Eq.~(\ref{eqn:dd}) which vanishes identically 
for $J_1 = J_2$, in agreement with the infinite degeneracy of the 
classical groundstate for $J_1<J_2$.
We already know, however, that the excitation spectrum of the DE 
model is different from that of a ferromagnet and, depending on the 
density of electrons, is softened either at the zone boundary or at 
$q=0$. 

A way to understand what this implies is to analyze what
form of the spin wave dispersion one generally 
expects at a point where a ferromagnet becomes unstable towards 
either canted or spiral phases, if there were no
degeneracy.   The simplest way to eliminate a degeneracy is to 
formally add an exchange interaction between second neighbors 
\be 
H_3 &=& J_3 \sum_{\langle ij \rangle^{\prime}} \vec{S}_i \vec{S}_j
\en
This extra interaction modifies the dispersion Eq.~(\ref{eqn:dd})
to give
\be
\omega_{sw}(q) &=& zS  (J_1 - J_2) (1 - \gamma_q) 
  - zS J_3 (1 - \gamma_{2q})
\en
For positive $J_3$ (antiferromagnetic next--nearest neighbour 
exchange), the $J_3$ term favors maximum misalignment of the 
next--nearest neighbour spins.  This condition is best satisfied 
by the spiral state.  
For negative $J_3$, the extra coupling favors parallel orientation 
of next--nearest neighbour spins.   This is a property of the canted 
phase. 
Analyzing the dispersion in the presence of $J_3$ we immediately see 
that for positive $J_3$, the spectrum first becomes unstable at 
$q=0$, at $J_1 = J_2 + 4 J_3$, while for negative $J_3$, the excitation 
spectrum first softens at $J_1 = J_2$ at $q = (\pi,\pi,\pi)$. 
Performing the classical spin wave analysis for the 
full DE Hamiltonian with extra $J_2$ and $J_3$ terms 
slightly below the instability we find that for positive $J_3$, 
the spin wave spectrum has three zero modes: 
at $q=0$ and  an incommensurate momenta ${\vec q} = \pm (q_0, q_0, 
q_0)$ where $q_0 = \theta$. This is exactly what we expect in a spiral 
state.  For negative $J_3$, the soft modes remain at $q=0$ and at 
$(\pi,\pi,\pi)$.  This is what we expect in a two-sublattice canted 
phase.

The above analysis shows that the type of the ground state 
configuration below the instability can be obtained by just analyzing 
the spin wave spectrum at the instability point of the ferromagnetic 
phase.  Using the results from the previous section we find that at $T=0$,
the instability occurs when 
\be
\tilde{\omega}_{sw}(q) &=& 
   zS(J_1 -J_2)[1-\gamma(q)] + \Delta\Sigma^{(2)}(q, 0) = 0
\en
where, as shown, $J_1 = \overline{t}/2S^2$ with $\overline{t} = t 
N^{-1}\sum_k n_k \gamma_k$, and the relevant self--energy 
correction is presented in Eq.~(\ref{T=0}). 
As discussed in Section \ref{DEFM}, this can be split into 
two of two pieces.  One scales as $1-\gamma(q)$ and just 
renormalize $J_1$.  The second piece $\Delta\Sigma^{(2)}(\vec{q})$ 
accounts for the renormalization of the form of the dispersion.
It reads 
\be 
\Delta\Sigma^{(2)}(\vec{q}) 
   &=& 4 \frac{zt}{(4S)^2} 
     \frac{1}{N^2} \sum_{12} \frac{\gamma_1^2 - \gamma_{1+q}^2
     }{\gamma_1 - \gamma_2} n_1(1-n_2)
\en
The first instability of the DEFM will occur for some value of 
$J_2$ sufficient to cancel the bare spin dispersion {\it and} 
this quantum correction.   The value of critical $J_2$ depends 
on the form of $\Delta\Sigma^{(2)}$ as a function of $q$ 
and doping $x$.  We therefore consider the ratio
\be
\lambda_q^{x} = \frac{\Delta\Sigma^{(2)}(q)}{1-\gamma(q)} 
\en 
as a function of doping.  The wavevector $q=q^{*}$ for 
which $\lambda_q^{x  }$ has its minimum value defines the 
momentum of the instability.

Near $q=0$, we obtain using (\ref{tt}) and (\ref{ttt})
\be
\lambda_{q\approx 0}^{x} 
   = \frac{t}{2S^2}~\frac{I(x)q^2}{q^2/z}  = \frac{zt}{2S^2}I(x) 
\en
while at the zone corner
\be
\lambda_{q\approx \pi}^{x} 
   = \frac{t}{2S^2}~\frac{I(x)q^2}{2-(\pi-q)^2/z} 
\to 0 
\en
Whether the first instability occurs in the zone center or the zone
corner therefore depends on the sign of $I (x)$.  
As we discussed in Sec~\ref{DEFM}),  $I (x)$ is positive for 
$x<0.31$ in both 2D and 3D, and for small density of 
holes $1-x < 0.07$ in $2D$ and  $1-x < 0.06$ in $3D$.
For these dopings, the intermediate configuration  is 
then a canted phase.  For intermediate densities
$0.31(7) < x < 0.92(7)$ in $2D$, and $0.31(7) < x < 0.94(2)$ in $3D$,
$I(x)$ is negative, \ie quantum effects cause a relative softening 
of spin wave modes at $q=0$, and the intermediate configuration is a 
spiral phase.  These results are in good agreement with earlier 
numerical studies studies~\cite{t=0}.

At finite temperature, the range of dopings where the canted phase is 
selected increases with $T$ as for realistic $x$, classical 
fluctuations primarily soften the dispersion near $q=0$ and hence 
favor the spiral state, at least at small $T$.  
At higher $T$,  and in $D=3$, thermal fluctuations may in principle, 
cause a re--entrant transition into a spiral phase as when $T$ becomes    
larger than $J_HS p^2_F$, the sign of $\Sigma^{(2)}_T (q)$ changes,
and thermal fluctuations now favor the canted phase.  
In practice, however, we found that this sign change occurs at 
unrealistically large $T$ for which our low $T$ analysis is 
inapplicable.
 
We conclude this section with few words of caution. 
First, in our analysis we assumed that there are no nontrivial 
transitions inside the intermediate phase, \ie the configuration 
which becomes the ground state immediately after ferromagnetic 
instability yields to another configuration at some distance away 
from the instability.  Such a possibility exists in general, but to 
the best of our knowledge, there is no known example of such behavior.
Second, we neglected phase separation~\cite{moreo}. 
To study this possibility in our approach, one has to analyze the 
sign of the longitudinal susceptibility in, e.g. the canted phase.  
If it is negative, then the system is unstable towards phase 
separation~\cite{ps}.  These calculations are currently under way.

\section{Conclusions}
\label{conclusions}

We have performed a novel large $S$ expansion for the Kondo model 
on a magnetically ordered lattice, and presented a simple,
physically transparent and controlled calculation scheme 
for evaluating corrections to magnetic and electronic 
properties of the model to ${\mc O}(1/S^2)$.
Calculations are particularly straightforward in the limit where the 
onsite (ferromagnetic) Kondo coupling $J_H$ greatly exceeds the electron 
bandwidth $zt$ (the double exchange model).
In this limit we have used our expansion scheme to study quantum 
and thermal corrections to the spin wave spectrum of the double 
exchange ferromagnet.
We argue that a double exchange ferromagnet is equivalent to a 
Heisenberg model with effective coupling 
$J_1 \propto t/S^2$ {\it only} at a quasi--classical
level --- both quantum and thermal corrections to the spin wave 
spectrum of a double exchange ferromagnet differ from {\it any} 
effective Heisenberg model because its spin excitations interact 
only indirectly, through the exchange of charge fluctuations. 
We demonstrated that the effective interaction between spin 
waves is mediated by Pauli susceptibility of electrons 
which has a much larger amplitude than $J_1$ and also has 
its own non--analytic momentum and frequency dependence. 
We argued that at $T=0$, the nonanalyticity in the frequency dependence 
of the effective vertex gives rise to a finite self-energy correction 
which changes the momentum dependence of the spin wave
dispersion  compared to that in a Heisenberg ferromagnet. 
The form of the full dispersion which we found is similar to that
found in numerical studies of the DE model 
\cite{kaplan}.  For a small density of electrons, or small density of holes, 
the corrections to dispersion lead to a ``softening'' of modes
at the zone boundary relative to those the zone center. 
For intermediate densities, quantum effects cause a relative softening 
of spin wave modes at $q=0$. These results are in good agreement with earlier 
numerical studies ~\cite{t=0}. 

The softening of zone boundary  spin waves has been observed in 
neutron scattering experiments on the CMR Manganites \cite{neutrons} 
at $x \sim 0.7$. 
It has been suggested that this softening may be due to deviations 
from Heisenberg behavior in the DEFM~\cite{kaplan,neutrons}, 
the influence of optical phonons \cite{furukawa99}, or due to orbital
degrees of freedom \cite{khaliullin}.  Our analysis argues against 
 the first possibility as for $x \sim 0.7$,  our zero
temperature theory predicts a relative hardening rather than softening
of the dispersion at $(\pi,\pi,\pi)$.  We note however that the leading
corrections to spin wave dispersion at finite $t/J_H$ {\it do} lead to
a softening of zone boundary modes \cite{natasha}.  

We also found that the process by which spin waves decay into 
lower energy spin excitations dressed with particle--hole pairs
leads to a finite spin wave lifetime at $T=0$.
The spin wave lifetime is very long at low $q$, scaling as $q^{-4}$ in 3D
for $p_F > q \to 0$, but increases rapidly with increasing $q$. 
The large absolute value of the damping at large $q$ in the DEFM 
and its strong momentum dependence is consistent with the 
experimental behavior in the Manganites \cite{neutrons}.

We then analyzed in detail the form of the temperature corrections to the 
spin wavespin wave dispersion.  We argued that for experimentally relevant densities, 
the corrections roughly have the same functional form as in a Heisenberg 
ferromagnet, but the overall scale is {\it much larger}.  At low density 
(small $p_F$), for which an analytic treatment is possible,
thermal corrections in DEFM are enhanced relative to those 
in a HFM with the same effective exchange coupling $J_1$ by a factor 
proportional to $1/p^2_F$.
At very small $p_F$, we also found that not only the amplitude but also 
the functional form of the thermal self-energy in DEFM is qualitatively 
different from that in a HFM.  In 3D, the thermal self-energy in DEFM 
at low doping even changes sign relative to that in a HFM.
The enhancement of temperature the overall scale corrections relative to a HFM
has been observed experimentally in La$_0.85$Sr$_0.15$MnO$_3$~\cite{lynn}, 
and we believe that our finite temperature results will be useful for the 
interpretation of the neutron data on Manganites.

We also considered the effect of a direct superexchange 
antiferromagnetic interaction between core spins. We find 
that the  competition between ferromagnetic double 
exchange and an antiferromagnetic superexchange  provides a new 
example of an ``order from disorder'' phenomenon ---  
when the two interactions are of comparable strength, 
an intermediate spin configuration (either a canted or a spiral 
state) is selected only by  quantum and/or thermal fluctuations. 
We discussed which configurations are selected at various dopings. 

The issue left for further study is the possibility of a
phase separation in the intermediate regime into ferromagnetic
and antiferromagnetic regions~\cite{numerical,maxim},
and also possible stripe formation at low hole doping. 
The analysis of these issues is clearly called for.
 
{\it Acknowledgments.}

It is our pleasure to acknowledge helpful conversation with 
S.\ W.\ Cheong, D.\ Golosov, R.\ Joynt,  M.\ Kagan, T.\ A.\ Kaplan, 
G.\ Khaliullin, S.\ D.\ Mahanti, A.\ Millis,  E.\ M{\"u}ller-Hartmann, 
M.\ Norman, O.\ Tchernyshyov, N.\ B.\ Perkins, N.\ M.\ Plakida 
and M.\ Rzchowski.
This work was supported under NSF grant DMR--9632527 
and the visitors program of MPI--PKS (N.\ S.), and by NSF 
grant DMR-9979749 (A.\ V.\ C.).


\begin{thebibliography}{}
\bibitem{comm1} 
   This is indeed true for the computations of the transverse 
   spin correlator between different sites.  For a Kondo problem at a 
   given site, higher--order terms in  the Holstein--Primakoff expansion 
   for a composite spin are indeed relevant as they preserve the correct 
   spin algebra, and are incorporated into our transformation in 
   Sec.~\protect\ref{transformation}.
\end{thebibliography}
\end{document}